\documentclass[a4paper]{article}

\usepackage{multirow}
\usepackage{caption}
\usepackage{INTERSPEECH2020}

\title{Intra-Utterance Similarity Preserving Knowledge Distillation for Audio Tagging}
\name{Chun-Chieh Chang$^1$ \footnotemark, Chieh-Chi Kao$^2$, Ming Sun$^2$, Chao Wang$^2$}
\address{
  $^{1}$Center for Language and Speech Processing, Johns Hopkins University. \\
  $^{1,2}$Alexa Speech, Amazon.com Inc.}
\email{cchunch1@jhu.edu, \{chiehchi, mingsun, wngcha\}@amazon.com}

\begin{document}

\maketitle
\footnotetext[1]{Work done while at Amazon.com Inc.}
\begin{abstract}
Knowledge Distillation (KD) is a popular area of research for reducing the size of large models while still maintaining good performance. The outputs of larger teacher models are used to guide the training of smaller student models. Given the repetitive nature of acoustic events, we propose to leverage this information to regulate the KD training for Audio Tagging. This novel KD method, “Intra-Utterance Similarity Preserving KD” (IUSP), shows promising results for the audio tagging task. It is motivated by the previously published KD method: “Similarity Preserving KD” (SP). However, instead of preserving the pairwise similarities between inputs within a mini-batch, our method preserves the pairwise similarities between the frames of a single input utterance. Our proposed KD method, IUSP, shows consistent improvements over SP across student models of different sizes on the DCASE 2019 Task 5 dataset for audio tagging. There is a $27.1\%$ to $122.4\%$ percent increase in improvement of micro AUPRC over the baseline relative to SP’s improvement of over the baseline. 
\end{abstract}


\section{Introduction}
In recent years, the release of commercial products such as Amazon Echo and Google Home has highlighted the need for compact neural network models. These devices offer a wide range of services that involve interacting with the user based on audio. The neural network models need to be small enough to fit on the device and yet accurate enough to be commercially viable. These smart devices have limited CPU and memory so the number of Floating Point Operations (FLOPs) and the number of model parameters are of great concern.

Knowledge Distillation (KD) is one popular technique used to improve results of smaller models. It works by having a large teacher model guide the training of a smaller student model. One of the first KD methods proposed uses logits from teacher and student models as an additional loss function \cite{1503.02531}. The output logits of the teacher can be viewed as soft targets for the student to achieve. This type of guided learning is not just limited to comparing outputs. There are also methods that compare the intermediate output features from selected layers within the teacher and student model. FitNet uses the outputs from an intermediate layer of the neural network and computes the Mean Squared Error between the intermediate outputs of the teacher model and the student model \cite{1412.6550}. There are numerous other KD methods but idea is that the teacher model guides the student model through an additional loss function based on a comparison between the two models \cite{1910.10699} \cite{Yim_2017_CVPR} \cite{1907.09682} \cite{Zagoruyko2017PayingMA} \cite{Park_2019_CVPR}. Many of these KD methods are published using Image Classifications datasets like CIFAR-10 or CIFAR-100 \cite{Krizhevsky09learningmultiple}.

There have been previous studies of KD for acoustic related tasks such as Speech Recognition \cite{Chebotar+2016} \cite{Kim2018BridgenetsST} \cite{Lu2017KnowledgeDF} \cite{li2014learning}, Acoustic Event Detection \cite{Shi2019CompressionOA} \cite{1910.11789} \cite{8683710}, and Acoustic Scene Classification \cite{Jung2019} \cite{Heo2019AcousticSC}. Our novel method, Intra-Utterance Similarity Preserving (IUSP) KD, attempts to leverage prior knowledge about the acoustic events when performing KD. In the Audio Tagging, there is often a lot of repetition that can be seen in the spectrogram of the audio clips. This is most apparent when viewing the spectrograms of stationary signals such as `car alarms' and `sirens'. For example, Figure 1 is the spectrogram of a siren from the DCASE 2019 Task 5 challenge \cite{Bello2019sonyc}.

\begin{figure}[htbp]
\begin{center}
	\includegraphics[width=50mm]{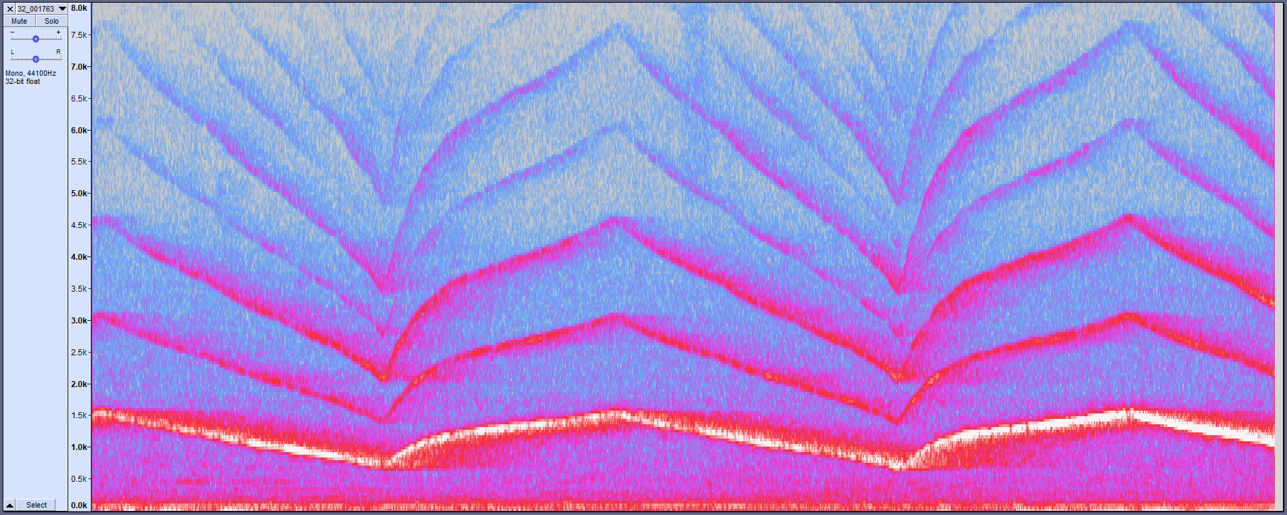} \includegraphics[width=20mm]{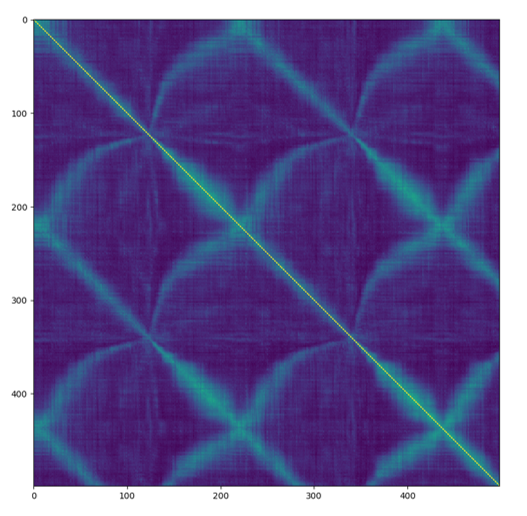}
	\caption{Left: Example spectrogram of audio clip for siren. Right: Intra-Utterance similarity matrix of siren clip.}
\end{center}
\end{figure}

When computing the pairwise similarity between frames, the resultant matrix should have strong values on the off-diagonals. This is seen in the Figure 1. Our proposed method compares the resultant similarity matrices from the intermediate features of both the teacher and student models. This ensures that the student model also captures information about the repetitive nature of the acoustic event.

The rest of the paper will be as follows: Section 2, an explanation of both Similarity Preserving KD (SP) and Intra-Utterance Similarity Preserving KD (IUSP); Section 3, a description of the audio tagging dataset used; Section 4, an overview of the experimental setup; Section 5, a presentation of the results; and Section 6, the conclusion.

\section{Method Description}
Section 2.1 describes the Similarity Preserving KD (SP) from literature \cite{1907.09682} which inspired our Intra-Utterance Similarity Preserving KD (IUSP) described in Section 2.2.
\subsection{Similarity Preserving KD}
Using the Audio Tagging task as an example, the principle behind Similarity Preserving KD \cite{1907.09682} is that inputs with the same event tags should have similar activations in the layers of the neural network. Given an intermediate output $A^{(l)} \in R^{b \times c \times h \times w}$, define $Q^{(l)} \in R^{b \times chw}$ as a reshaping of $A^{(l)}$. Where $l$ is the layer number, $b$ is the batch size, $c$ is the output channels, and $h$ and $w$ are the input dimensions. In our task, $h$ is related to number of log mel-frequency bins and $w$ is related to the number of frames. Note that various layers may reduce $h$ and $w$ so they may not have the exact same values as the original input dimensions.

The pairwise similarities between each audio clip in the batch can be computed using the following equations:
\begin{equation}
\tilde{G}^{(l)} = Q^{(l)} \cdot Q^{(l)\top}; \;
G^{(l)}_{[i,:]} = \tilde{G}^{(l)}_{[i,:]} / ||\tilde{G}^{(l)}_{[i,:]}||_2
\end{equation}
Where $i$ denotes the row and so $G^{(l)}$ is the row-wise normalized version of $\tilde{G}^{(l)}$. Since the goal of Similarity Preserving KD is to ensure that the student learns the same pairwise similarity matrix as the teacher, this matrix $G^{(l)}$ is computed for both the teacher and the student models. The Similarity Preserving KD loss is thus defined:
\begin{equation}
\mathcal{L}_{SP} = \frac{1}{b^2} \sum_{(l,l') \in \mathcal{I}} ||G^{(l)}_{Teacher} - G^{(l')}_{Student}||^{2}_F
\end{equation}
Various different layers of the student and teacher model can be compared against each other so $\mathcal{I}$ is the collection of the layer pairs. With $l$ being the layer index of the teacher model and $l'$ being the layer index of the student model.

\subsection{Intra-Utterance Similarity Preserving KD}
As mentioned in the Introduction, our proposed method is based on Similarity Preserving KD \cite{1907.09682}. However, instead of preserving the pairwise similarities between utterances in a batch, we preserve the pairwise similarities between frames of an utterance.

Given an intermediate output $A^{(l)} \in R^{b \times c \times h \times w}$ as defined in Section 2.1 above. Since $A^{(l)}$ includes the entire batch, call the integer $b' \in [1,b]$ the index of a specific utterance in the batch. First, normalize $A^{(l)}$ along the channel dimension using an equation shown below. This way no channel has greater weight than the others. Then, compute the similarity matrix between frames, also shown below.
\begin{equation}
G_{(b')} = Q^{(l)}_{[b']} \cdot Q^{(l)\top}_{[b']}; \;
\tilde{A}^{(l)}_{[b',i,:,:]} = A^{(l)}_{[b',i,:,:]}/||A^{(l)}_{[b',i,:,:]}||_2
\end{equation}
$Q^{(l)}_{[b']} \in R^{ch \times w}$ is defined as a reshaped version of $\tilde{A}^{(l)}_{[b',:,:,:]} \in R^{c \times h \times w}$. The matrix $G_{(b')}$ is the pairwise similarity between each frame of utterance $b'$ in the batch.
This similarity matrix is computed for both the student and teacher model. In the event that $h$ and $w$ from the teacher model are different than those of the student model, bilinear interpolation is used to make the dimensions of $A^{(l)}_{Teacher}$ match $A^{(l)}_{Student}$. The Intra-Utterance Similarity Preserving KD loss function thus is defined:
\begin{equation}
\mathcal{L}_{IUSP} = \frac{1}{b} \sum_{b' \in [1,b]} || G^{Teacher}_{(b')} - G^{Student}_{(b')} ||^{2}_{F}
\end{equation}
One modification made to the loss function above was the addition of a sigmoid function applied to each element in matrix $G_{(b')}$. This ensures that the differences between high and low similarities are exaggerated. The final loss function is defined as: 
\begin{equation}
\mathcal{L}_{IUSP} = \frac{1}{b} \sum_{b' \in [1,b]} || \tilde{G}^{Teacher}_{(b')} - \tilde{G}^{Student}_{(b')} ||^{2}_{F}
\end{equation}
Where:
\begin{equation}
\tilde{G}_{(b')} = sig(\gamma \times (G_{(b')} - \delta))
\end{equation}
The hyper parameters $\gamma$ and $\delta$ are meant to scale and shift the sigmoid function so that it can serve as a threshold for determining high and low similarity. We selected the values $10$ and $0.5$ for $\gamma$ and $\delta$ respectively after tuning on the validation data of the dataset we used.

\begin{figure}[htbp]
\begin{center}
	\includegraphics[width=80mm]{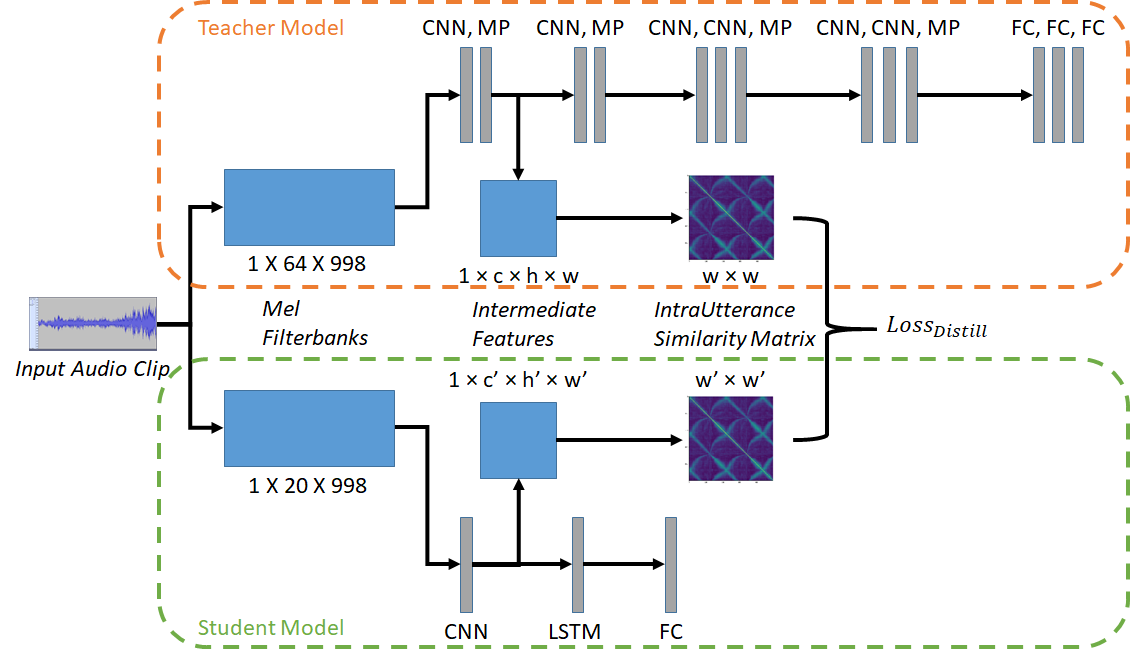}
	\caption{IntraUtterance Similarity Preserving KD. MP is Max Pooling and FC is Fully Connected Layer. The pairwise similarity of each frame in a given input of a batch is computed for both the teacher and student network. These matrices are then compared against each other. This IntraUtterance Similarity Preserving KD can be used alongside other KD methods.}
\end{center}
\end{figure}

\section{Dataset Description}
The dataset used is the DCASE 2019 Task 5 dataset \cite{Bello2019sonyc}. It is an Audio Tagging task where the goal is to predict whether or not any of the 8 coarse-grained classes are present in the 10s audio clip. The given development data is split into 2351 clips of training data and 443 clips of validation data. The 274 clips of test data was released after the competition. The metric used for this competition was Area Under Precision and Recall Curve (AUPRC). The higher the AUPRC the better. The detection threshold was incrementally raised and the global tally of true positives (TP), false positives (FP), and false negatives (FN) at that threshold was computed by summing together the individual tallies of TP, FP, and FN of each category for the same threshold. This global TP, FP, and FN at each threshold is then used to calculate the precision recall. These precision recall values are then plotted and the trapezoidal rule is used to compute the Area Under the Precision Recall Curve. This AUPRC was used as the evaluation metric for the competition and our experiments. All the evaluations for our setup were done using the code provided by the competition organizers and can be found on github \cite{Bello2019sonyc}.

\section{Experimental Setup}
\subsection{Models}
The teacher model used in our experiments is the second place model of the DCASE 2019 Task 5 challenge \cite{Kim2019}. We received the model weights from the author directly. This model used pretrained weights from the first six layers of the VGGish model \cite{Hershey2017CNNAF} as a starting point for transfer learning. The AUPRC of $0.837$ for the second place model was good enough for the purposes of our experiments. This model was chosen because the first place model had additional feature augmentation techniques that were not directly relevant to experiments of our novel KD method.

The student model we used is a CNN-LSTM model. The student model has one CNN layer, one LSTM layer, and one fully connected layer. Table 1 shows the model architecture of the student model in more detail. This design choice was motivated by the need to keep the student model as small as possible while still keeping in line with what is conventionally used in Audio Tagging. The first CNN layer was chosen because the top teams in the DCASE 2019 Task 5 challenge use CNN based architectures \cite{Adapa2019}\cite{Kim2019}\cite{Cui2019}. In addition, CNNs have been shown to work well in audio tasks \cite{Hershey2017CNNAF} \cite{Takahashi2016DeepCN}. The LSTM layer was chosen because LSTMs are suited for capturing time series information. More generally, RNNs have had success with audio tasks as well \cite{8461975} \cite{Parascandolo2016RecurrentNN}. The hidden dimensions of the LSTM layer in the student model were varied from 128 hidden dimensions to 16 hidden dimensions to test the robustness of the proposed method on different sizes of small models. Table 2 shows the number of parameters for the teacher and student models.

\begin{table}[htbp]
	\caption{Architecture of CNN-LSTM Model Used}
	\begin{center}
		\begin{tabular}{clc}
			\hline \hline
			Layer & Type & Configuration\\
			\hline
			1 & CNN & num\_filters: 32 filter\_size: 5x5 stride: 2\\
			2 & LSTM & hidden\_dim: 128, 64, 32, or 16\\
			3 & FC & \\
			\hline \hline
		\end{tabular}
	\end{center}
\end{table}

\begin{table}[htbp]
	\caption{Number of FLOPS and Parameters in Student and Teacher Models}
	\begin{center}
		\begin{tabular}{llcc}
		\hline \hline
		& & FLOPS (G) & Params (M) \\
		\hline
		Teacher & & 14.7 & 3.96 \\
		\hline
		Student & LSTM 128 & 0.33 & 0.742 \\
		& LSTM 64 & 0.17 & 0.355 \\
		& LSTM 32 & 0.08 & 0.173 \\
		& LSTM 16 & 0.04 & 0.086 \\
		\hline \hline
		\end{tabular}
	\end{center}
\end{table}

\subsection{Feature Extraction}
We followed the same feature extraction process as the second place winner of DCASE 2019 Task 5, matching our teacher model selection. First the audio clips were resampled to $16kHz$. Then log mel-filterbanks were computed with a window of $25ms$ and step size of $10ms$. Each sample has 64 bins so the final feature dimension input to the teacher model is $64 \times 998$. Keeping in line with the theme of making the student model as small as possible, the student model features uses 20 bins so the feature dimensions for the student model is $20 \times 998$.

\subsection{Experiments}
There were five setups for our experiments with different $\mathcal{L}_{Total}$ used as the objective function for training: Baseline `BCE', `BCE+KD', `BCE+KD+SP', `BCE+KD+IUSP', and Combination `BCE+KD+SP+IUSP'. Where `BCE' is Binary Cross Entropy, `KD' is the original KD paper \cite{1503.02531}, `SP' is the Similarity Preserving KD \cite{1907.09682}, and `IUSP' is our proposed method Intra-Utterance Similarity Preserving KD. An example total loss function is shown below:
\begin{equation}
\mathcal{L}_{Total} = \alpha_1 \mathcal{L}_{BCE} + \alpha_2 \mathcal{L}_{KD} + \alpha_3 \mathcal{L}_{SP} + \alpha_4 \mathcal{L}_{IUSP}
\end{equation}


For all experimental setups, the $\alpha$ hyper parameters were chosen so that all the loss items, $\alpha_1 \mathcal{L}_{BCE}$; $\alpha_2 \mathcal{L}_{KD}$; $\alpha_3 \mathcal{L}_{SP}$; and $\alpha_4 \mathcal{L}_{IUSP}$, were of equal magnitude. The values for $\alpha_{1...4}$ are: 1.0, 10.0, 10.0, and 1.0 respectively. If a particular setup did not include a specific loss item, then the $\alpha$ value for that loss item was $0.0$.


The student models were trained using the Adam Optimizer with a learning rate of $0.0001$ for 300 epochs. There was also a early stopping criteria where training stops if the AUPRC of the validation set did not improve for more than 20 epochs. In practice, most of the training plateaued around 50 epochs.

\subsection{Tuning Intermediate Hint Layers}
Given that the teacher and student models have significantly different architectures, there are many different choices when choosing potential intermediate outputs to compare. Referring to Figure 2, for the teacher model, we used the outputs after the Max Pooling layers as potential hint layers to guide the student model. For the student model, there were only three layers in total so there are only two possible intermediate hint layers to use. Four potential hint layers from the teacher model and two potential hint layers from the student model means that we tried eight different combinations of $(A^{(l)}_{Teacher}, A^{(l)}_{Student})$ pairs that can be used to calculate the SP and IUSP losses.

Four trials each were performed on the validation dataset with all possible combinations and for all LSTM sizes. The configuration with the best average micro AUPRC across all LSTM sizes was chosen for final results and analysis. For `BCE+KD+SP', we use the intermediate output from the second Max Pooling layer of the teacher and the intermediate output from the CNN layer of the student. For `BCE+KD+IUSP', we use the intermediate output from the first Max Pooling layer of the teacher and the intermediate output from the CNN layer of the student. For `BCE+KD+SP+IUSP', both `SP' and `IUSP' used the intermediate output from the second Max Pooling layer of the teacher and the intermediate output of the CNN layer from the student.

\section{Results and Analysis}
\subsection{Top Level Results}
\begin{figure}[htbp]
	\centering
	\includegraphics[width=75mm]{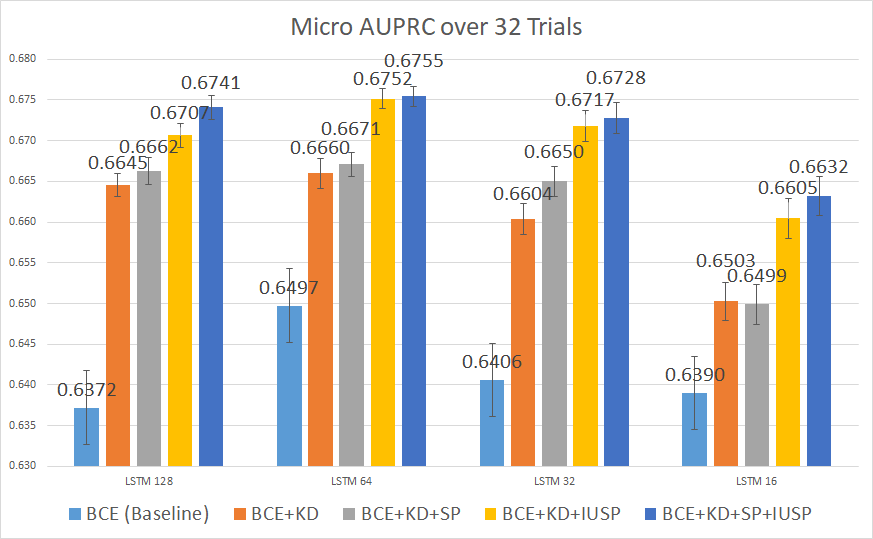}
	\caption{Overall Micro AUPRC for each Student Model with standard error bars. The higher the AUPRC the better.}
\end{figure}

A full graph of the micro AUPRC results is shown in Figure 3 with standard error of the mean computed from 32 trials of each experimental setup. As mentioned before, the teacher model has a micro AUPRC of $0.837$. In our experiments, the dark blue bar, the setup `BCE+KD+SP+IUSP', has the highest micro AUPRC over 32 trials across all different model sizes. The relative improvements over the Baseline and `BCE+KD+SP' is pretty consistent throughout all the varying LSTM dimensions. It seems that there is an overall trend that adding `SP' to the experimental setups improves results. However, any improvement is small and the standard error for setups with and without `SP' overlaps.

Table 3 shows the relative and absolute improvements. The addition of our KD method, `BCE+KD+SP+IUSP', provides a 27.1\% to 122.4\% increase in performance compared to Similarity Preserving KD, `BCE+KD+SP'.

\begin{table}[htbp]
	\caption{Relative and Absolute Improvements of Various KD Methods for different LSTM hidden dimensions.}
	\begin{center}
		\begin{tabular}{lcccc}
			\hline \hline
			& \textbf{128} & \textbf{64} & \textbf{32} & \textbf{16} \\
			\hline
			\textbf{Micro AUPRC} &&&& \\
			\hline
			BCE & 0.637 & 0.650 & 0.641 & 0.639 \\
			BCE+KD+SP & 0.666 & 0.667 & 0.665 & 0.650 \\
			{\begin{footnotesize}BCE+KD+SP+IUSP\end{footnotesize}} & \textbf{0.674} & \textbf{0.675} & \textbf{0.673} & \textbf{0.663} \\
			\hline
			\multicolumn{3}{l}{\textbf{Improvement over Baseline BCE}} && \\
			\hline
			BCE+KD+SP & 0.029 & 0.017 & 0.024 & 0.011 \\
			{\begin{footnotesize}BCE+KD+SP+IUSP\end{footnotesize}} & 0.037 & 0.026 & 0.032 & 0.024 \\
			\hline
			Percent Increase &27.1\%&48.2\%&31.9\%&122.4\% \\
			\hline \hline
		\end{tabular}
	\end{center}
\end{table}

\subsection{Class Level AUPRC Results}
Table 4 is of the class-wise results for the models with LSTM 128. The full results for all five setups are not shown because of space constraints.

\begin{table}[htbp]
	\caption{Class-wise AUPRC. Setup 1 is `BCE+KD', Setup 2 is `BCE+KD+SP', Setup 3 is `BCE+KD+IUSP', and Setup 4 is `BCE+KD+SP+IUSP'}
	\begin{center}
		\begin{tabular}{lcccc}
			\hline \hline
			& Setup 1 & Setup 2 & Setup 3 & Setup 4 \\
			\hline
			\textbf{LSTM 128} &&&& \\
			\hline
			\multirow{2}{*}{engine} & \multirow{2}{*}{0.7938} & \multirow{2}{*}{0.8038} & \multirow{2}{*}{0.8009} & \multirow{2}{*}{\textbf{0.8147}} \\
			&&&& \\
			machinery & \multirow{2}{*}{0.5081} & \multirow{2}{*}{0.5021} & \multirow{2}{*}{0.5606} & \multirow{2}{*}{\textbf{0.5659}} \\
			-impact &&&& \\
			non-machinery & \multirow{2}{*}{0.0941} & \multirow{2}{*}{0.0889} & \multirow{2}{*}{0.0955} & \multirow{2}{*}{\textbf{0.1071}} \\
			-impact &&&& \\
			\multirow{2}{*}{powered-saw} & \multirow{2}{*}{0.4394} & \multirow{2}{*}{0.4477} & \multirow{2}{*}{\textbf{0.4522}} & \multirow{2}{*}{0.4429} \\
			&&&& \\
			\multirow{2}{*}{alert-signal} & \multirow{2}{*}{0.6033} & \multirow{2}{*}{0.6164} & \multirow{2}{*}{0.6112} & \multirow{2}{*}{\textbf{0.6367}} \\
			&&&& \\
			\multirow{2}{*}{music} & \multirow{2}{*}{\textbf{0.1564}} & \multirow{2}{*}{0.1343} & \multirow{2}{*}{0.1532} & \multirow{2}{*}{0.1074} \\
			&&&& \\
			\multirow{2}{*}{human-voice} & \multirow{2}{*}{\textbf{0.7357}} & \multirow{2}{*}{0.7255} & \multirow{2}{*}{0.7280} & \multirow{2}{*}{0.7119} \\
			&&&& \\
			\multirow{2}{*}{dog} & \multirow{2}{*}{0.6422} & \multirow{2}{*}{\textbf{0.6438}} & \multirow{2}{*}{0.6360} & \multirow{2}{*}{0.6427} \\
			&&&& \\
			\hline \hline
		\end{tabular}
	\end{center}
\end{table}

The best method for each category is mostly consistent throughout all the different LSTM hidden dimensions. Figure 4 shows the difference in class-wise AUPRC of the combination `BCE+KD+SP+IUSP' system over the Baseline `BCE+KD' system for different LSTM hidden dimensions. There is improvement in some categories and degradation in others. The overall improvement in AUPRC is likely dataset dependent.

The improvements seen are in line with what is expected given the preservation of Intra-Utterance Similarity. Events like `machinery-impact' and `non-machinery-impact' are likely loud, singular sounds. Events like `alert-signal' are likely loud, repetitive sounds. Both of these types of events would have obvious indications in the spectrogram. So `BCE+KD+SP+IUSP' performs the best in all the aforementioned categories. 

\begin{figure}[htbp]
	\centering
	\includegraphics[width=75mm]{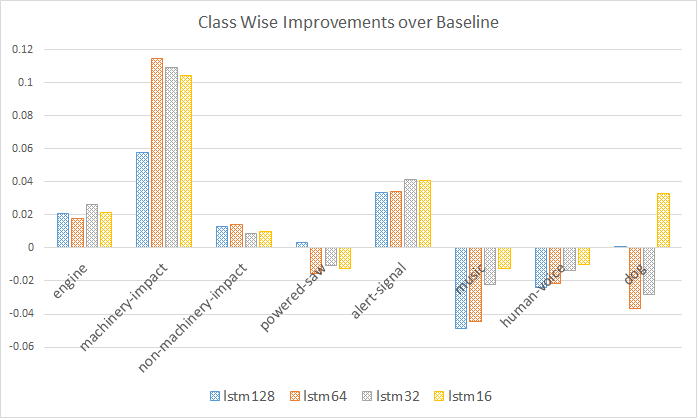}
	\caption{The class-wise improvements of the combination `BCE+KD+SP+IUSP' system over the baseline `BCE+KD' system. This is the difference in class-wise AUPRC.}
\end{figure}

Going back to the spectrogram and Intra-Utterance Similarity matrix of Figure 1, the pattern that is enforced is very clear. Things like `music' and `human-voice' are a lot more varied in terms of frequency in the spectrogram. So in general, there is no meaningful `IUSP' to be learned. An example spectrogram of the `human-voice' category is shown in Figure 5.

\begin{figure}[htbp]
	\begin{center}
		\includegraphics[width=50mm]{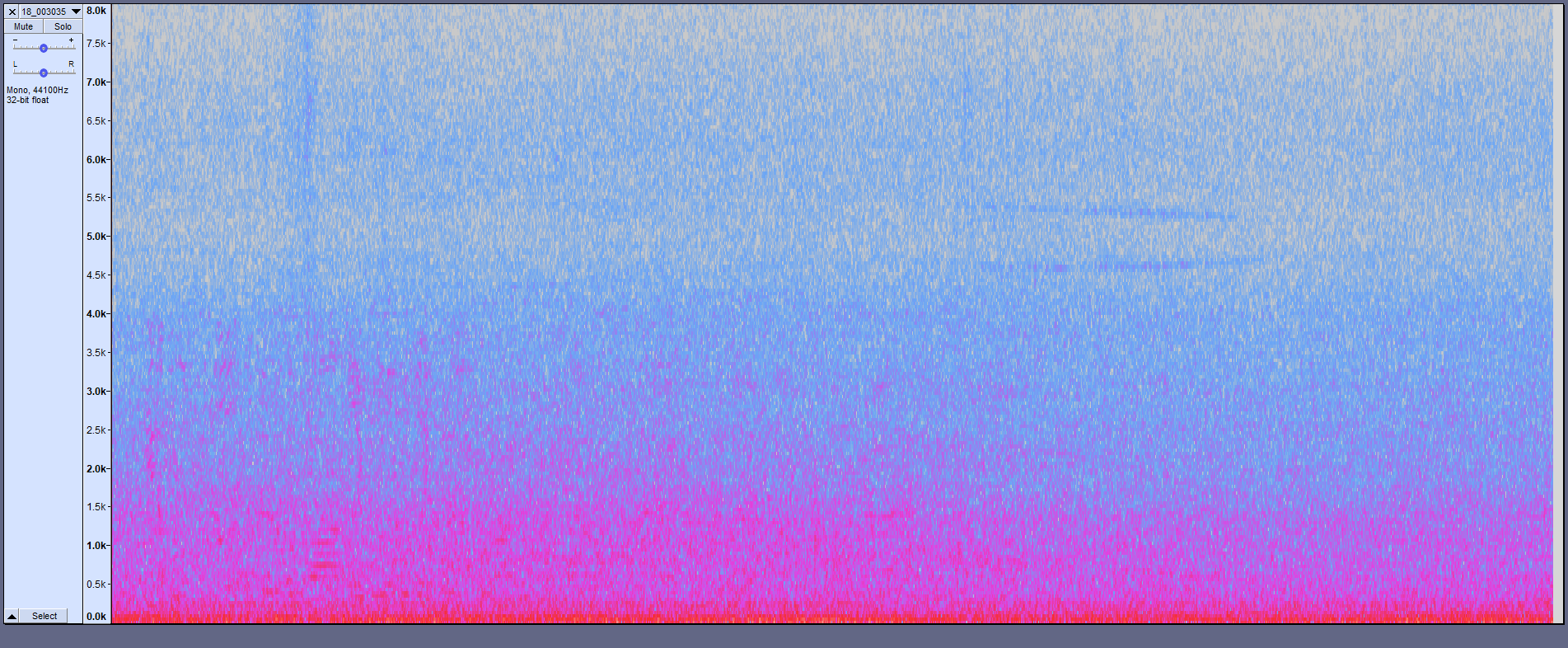} \includegraphics[width=20mm]{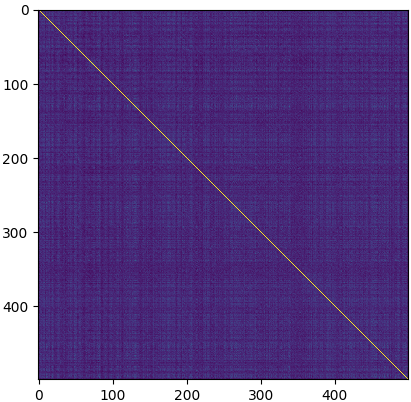}
		\caption{Spectrogram and `IUSP' Matrix for `human-voice' category.}
	\end{center}
\end{figure}

This hypothesis that our proposed method only significantly improves results for strong stationary signals or singular high energy events is bolstered by the fact that we did not see any improvements on Audio Scene Classification. We tried using our proposed method on the DCASE 2019 Task 1 \cite{Mesaros2018_DCASE} with a top team \cite{Koutini2019} as the teacher model. The different scene classes were of public areas such as `metro', `bus-station', and `shopping-mall'. The experimental setup and procedure used for Task 1 was the same as the one used for DCASE 2019 Task 5. However, when comparing the results there was no clear difference between all 5 setups. In terms of class-wise performance, there was also no clear pattern to the class-wise improvements or degradations.

\section{Conclusion}
In conclusion, our proposed KD method, Intra-Utterance Similarity Preserving KD, shows improvement over Similarity Preserving KD. This is true for both setups `BCE+KD+IUSP' and `BCE+KD+SP+IUSP', though the combination of all the loss items performs the best. There is a $27.1\%$ to $122.4\%$ percent increase in performance. Experimental results indicate that this method performs best when the dataset contains many audio clips of sounds with strong repetitions or loud singular events. Further experiments on a wider range of classes and datasets may yield a better heuristic in determining the usefulness of `IUSP'.

\bibliographystyle{IEEEtran}

\bibliography{mybib}

\end{document}